\documentclass[12pt]{article}

\usepackage{xcolor}
\usepackage[T1]{fontenc} 
\usepackage{amsmath,amsfonts,amsthm} 
\usepackage[utf8]{inputenc}
\usepackage{graphicx}
\usepackage{t1enc}
\usepackage{amssymb}

\usepackage{dcolumn}
\usepackage{bm}
\usepackage{float}
\usepackage{comment}
\usepackage{times}
\usepackage{multirow}
\usepackage[authoryear]{natbib}
\usepackage{rotating}
\usepackage{latexsym}
\usepackage{bbm}
\usepackage{bbold}
\usepackage{breakcites}
\usepackage{lineno}
\usepackage{ulem}
\usepackage{censor}

\definecolor{purpura}{rgb}{0.5, 0.0, 0.5}
\definecolor{azul}{rgb}{0, 0.0, 0.6}
\definecolor{rojo}{rgb}{0.6, 0, 0}
\definecolor{verde}{rgb}{0, 0.4, 0}
\definecolor{turquesa}{rgb}{0, 0.5, 0.5}
\definecolor{marron}{rgb}{0.6, 0.4, 0}
\definecolor{gris}{rgb}{0.4, 0.4, 0.4}
\definecolor{celeste}{rgb}{0.5, 0.5, 0.8}

\newcommand{\captionfonts}{\normalsize}

\makeatletter  
\long\def\@makecaption#1#2{%
  \vskip\abovecaptionskip
  \sbox\@tempboxa{{\captionfonts #1: #2}}%
  \ifdim \wd\@tempboxa >\hsize
    {\captionfonts #1: #2\par}
  \else
    \hbox to\hsize{\hfil\box\@tempboxa\hfil}%
  \fi
  \vskip\belowcaptionskip}
\makeatother   

\begin{document}

\hspace{13.9cm}1

\ \vspace{20mm}\\

{\bf \LARGE Statistical properties of color matching functions}

\ \\

{\bf \large Mar\'{\i}a da Fonseca$^{\displaystyle 1, \displaystyle 2}$ and In\'es Samengo$^{\displaystyle 1}$}

{$^{\displaystyle 1}$Instituto Balseiro, CONICET, and Department of Medical Physics, Centro at\'omico Bariloche, Argentina.}

{$^{\displaystyle 2}$Center for Brain and Cognition, and Department of Information and Communication Technologies, Universitat Pompeu Fabra, Barcelona, Spain.}
%



\thispagestyle{empty}
\markboth{}{NC instructions}
%
%

\begin{center} {\bf Abstract} \end{center}
In trichromats, color vision entails the projection of an infinite-dimensional space (the one containing all possible electromagnetic power spectra) onto the 3-dimensional space that modulates the activity of the three types of cones. This drastic reduction in dimensionality gives rise to metamerism, that is, the perceptual chromatic equivalence between two different light spectra. The classes of equivalence of metamerism are revealed by color-matching experiments, in which observers adjust the intensity of three monochromatic light beams of three pre-set wavelengths (the {\sl primaries}) to produce a mixture that is perceptually equal to a given monochromatic target stimulus. Here we use the linear relation between the color matching functions and the absorption probabilities of each type of cone to find particularly useful triplets of primaries. As a second goal, we also derive an analytical description of the trial-to-trial variability and the correlations of color matching functions stemming from Poissonian noise in photon capture. We analyze how the statistical properties of the responses to color-matching experiments vary with the retinal composition and the wavelengths of peak absorption probability, and compare them with experimental data on subject-to-subject variability obtained previously.

\newpage 

\renewcommand{\tablename}{Tabla}

\parskip 18pt
\baselineskip 0.3in


\section{Introduction}

Color vision has limitations. If we are instructed to provide objective measures of the percept produced by a chromatic stimulus, our responses are endowed with some degree of trial-to-trial variability, evidencing that the physical properties of the stimulus determine the subjective experience only up to a certain degree. In this paper, we provide an analytical derivation of the variability based on a probabilistic description of cone functioning. This variability has strong consequences in industrial applications involving the fabrication of computer screens, or chemical pigments. Quite unfortunately, the scientific communities working in colorimetry and computational neuroscience only seldom talk to each other.  This paper is an attempt to facilitate the dialogue between the two fellowships. 

Our goal is to provide an analytical description of the variability that stems from the stochasticity of photon absorption by cones. In particular, this stochasticity introduces noisy responses in color matching experiments. The magnitude, correlations, and wavelength dependence of the fluctuations of the responses are determined by the proportion of $\mathrm{s}, \mathrm{m}$ and $\ell$ cones of the retina of the observer, as well as the shape of the cone fundamentals, that is, the curves describing the absorption probability of photons of different wavelength by each type of cone. To describe these effects, we first review the mathematics of color matching experiments in Sect.~\ref{s:cme}. In passing, in Sect.~\ref{s:triplets} we discuss different strategies to select triplets of primaries that may be particularly convenient. We then move to Sect.~\ref{s:varcovar} to model the statistics of the trial-to-trial fluctuations in color matching experiments. In Sect.~\ref{derivation}, we use the Cr\'amer Rao bound on the Fisher information to derive an analytical expression for the variance and covariance of the fluctuations within an ideal-observer scheme. The dependence of the results on the physiological properties of the retina of the observer are discussed in Sect.~\ref{s:retinas}, and the triplet of primary colors that yield minimal trial-to-trial variability are derived in Sect.~\ref{s:primariosinformativos}. The lack of detailed experimental data on the within-subject fluctuations in color matching experiments does not allow us to test our theoretical predictions. Yet, in Sect.~\ref{s:experiments}, we compare the statistical properties predicted for within-observer fluctuations with those obtained from multiple observers. Since only a qualitative resemblance can be claimed, we believe that (as also discussed in Sect.~\ref{s:retinas}), the individual differences in the physiological properties of the visual system of different observers somewhat blurs the within-observer results.  We conclude the paper with a short summary of the main findings.


\section{Color matching experiments}
\label{s:cme}

In two previous papers \citep{Fonseca2016, Fonseca2018}, we showed that although there are many putative sources of variability in the visual pathway, the Poissonian nature of photon absorption by cones suffices to explain a large fraction of the variance in discrimination experiments \citep{MacAdam1942}. When a light beam of quantal distribution $I(\lambda)$ impinges on the retina, the three types of color-sensitive photoreceptors, cones of type $S$, $M$ and $L$ absorb $\bm{k}' = (k'_{{\mathrm s}}, k'_{{\mathrm m}}, k'_\ell)^t$ photons with probability distribution \citep{Zhaoping2011}
\begin{equation}
P[\bm{k}'|I(\lambda)] = \prod_{i \in \{ {\mathrm s}, {\mathrm m}, \ell\}} \textrm{Poisson} (k'_i|\alpha_i),
\label{Pr}
\end{equation}
where each Poisson factor reads
\[
\textrm{Poisson}(k'|\alpha) = {\rm e}^{-\alpha} \ \frac{\alpha^{k'}}{k'!},
\]
with mean and variance
\begin{equation}
\alpha_i = \beta_i \ \int I(\lambda) \ q_{i}(\lambda) \ \mathrm{d} \lambda, \ \ \ \ \ \ \ \ i \ \in \ \{{\mathrm s}, {\mathrm m}, \ell \}. \label{e1}
\end{equation}
Throughout the paper, primed quantities  vary from trial to trial. The parameters $\beta_i$ represent the fraction of each type of cone in the retina of the observer, and the curves $q_i(\lambda)$ are the cone fundamentals describing the wavelength dependence of the absorption probability of each type \citep{Stockman2010}. In Eq.~\ref{e1}, the space of all possible light spectra $I(\lambda)$ is projected on the $3$-dimensional space of vectors $\bm{k}'$. Importantly, the projection is probabilistic, and in different trials, the same spectrum $I(\lambda)$ may generate different $\bm{k}'$-vectors. The mean value of the number of absorbed photons of each type is $\langle \bm{k}' \rangle = \bm{\alpha} = (\alpha_{\mathrm{s}}, \alpha_{\mathrm{m}}, \alpha_\ell)^t$. 

Equation~\ref{e1} is not only an algorithm to calculate the mean and variance of the distribution of Eq.~\ref{Pr}, but also a linear projection of $I(\lambda)$ into the triplet $\bm{\alpha}$. Thus interpreted, Eq.~\ref{e1} provides the $LMS$ color coordinates \citep{Wyszecki1971}. The vector $\bm{\alpha}$ is therefore a 3-dimensional representation of $I(\lambda)$, in other words, it defines a possible linear space in which color can be specified. The components of $\bm{\alpha}$ depend on the observer, since both $\bm{\beta}$ and $\bm{q}(\lambda)$ vary up to a certain degree from subject to subject. However, they do not vary from trial-to-trial for the same observer. Instead, $\bm{k}'$ is a noisy instance of the representation, the one with which the visual system symbolizes the same spectrum in a single-trial. Importantly, this noisy signal is the only message that reaches the brain carrying chromatic information. 

\begin{figure}[ht]
\begin{center}
\includegraphics[scale=0.24]{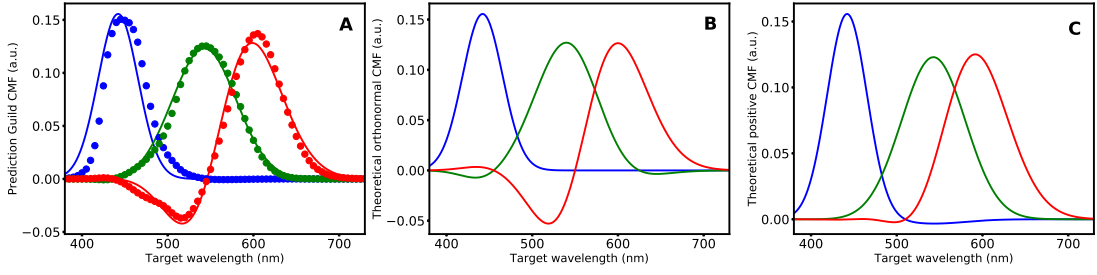}
\caption{A: Observer-averaged color matching functions reported by Guild \citep{Guild1932}, employed by the CIE 1931 to construct their RGB and XYZ color spaces ($\lambda_1 = 435.8$ nm in blue dots, $\lambda_2 = 546.1$ nm in green dots and $\lambda_3 = 700$ nm in red dots) normalized to unit Euclidean norm, and the corresponding normalized theoretical prediction in solid lines (Eq.~\ref{e:efes}). B: Predicted normalized CMF corresponding to primaries $\lambda_1 = 455$ nm (blue), $\lambda_2 = 550$ nm (green), $\lambda_3 = 625$ nm (red), selected to minimize the scalar product between the curves. C: Predicted normalized CMF corresponding to primaries $\lambda_1 = 380$ nm (blue), $\lambda_2 = 510$ nm (green), $\lambda_3 = 775$ nm (red), selected to maximize the range of $\lambda_{{\mathrm t}}$ values for which the curves are positive. In all panels, we constructed the matrix $Q$ with the cone fundamentals of \citet{Stockman2000}.} \label{CMF}
\end{center}
\end{figure}

It is interesting to note that long before the anatomy and the physiology of photoreceptors were described, Hermann von Helmholtz and Thomas Young predicted their existence from the $3$-dimensional nature of chromatic percepts demonstrated by phsychophysical experiments \citep{Helmholtz1910}. Their starting point was the fact that any chromatic sensation can be perceptually equated to a combination of three monochromatic beams of adjustable intensity, the so-called \textit{primary colors}. The triplet of primaries is not unique, since many choices can be used, as long as the mixture of the two colors does not produce the chromatic sensation of the third. In the $19^{\mathrm{th}}$ century, Hermann Grassmann \citep{Grassmann1853} introduced the laws that carry his name, and govern the rules of color matching: symmetry, transitivity, proportionality and additivity  \citep{Wyszecki2000}. In 1931, the Commission internationale de l'éclairage (CIE) reported the results for a collection of the experiments called \textit{color matching experiments} \citep{CIE1932}. Subjects were instructed to adjust the gains $g'_1, g'_2, g'_3$ of three monochromatic beams of wavelengths $\lambda_1, \lambda_2, \lambda_3$ and intensities $I_1, I_2, I_3$ (the primaries) to match a \textit{target} spectral color of wavelength $\lambda_{\mathrm{t}}$. The experimenter showed a bipartite field on a screen. One of the halves was illuminated with the \textit{target} stimulus, of spectrum 
\begin{equation} \label{e4}
I_{\mathrm{t}}(\lambda) = I_{\mathrm{t}} \ \delta(\lambda - \lambda_\mathrm{t}),
\end{equation}
and the other half displayed the \textit{matched} color, of spectrum
\begin{equation} \label{e5}
I_{\mathrm{m}}(\lambda) = g'_1 \ I_1 \ \delta(\lambda-\lambda_1) + g'_2 \ I_2 \ \delta(\lambda-\lambda_2) + g'_3 \ I_3 \ \delta(\lambda-\lambda_3). 
\end{equation}
In these equations, the primed quantities $g'_1, g'_2, g'_3$ are stochastic variables that represent the result of a given subject to a single instance of the matching experiment. In this paper, we define the trial average $g_i = \langle g'_i \rangle$ of the gains as the red, the green and the blue \textit{color matching functions} (CMF) of a given observer. This definition contrasts with the one employed by the CIE 1931, where they calculated a population average over the responses of 18 subjects, instead of a trial average. The so-called ``standard observer'' of Fig.~\ref{CMF}A \citep{Wyszecki2000} displays the result of such a population average, for a set of target wavelengths $\lambda_{\mathrm{t}} \in [\mathrm{380 \ nm, 780 \ nm}]$ every $5$ nm. If the averaged subjects all have the same retinal composition and the same absorption probabilities, the two definitions coincide.

For the time being, we restrict the analysis to trial averages for a single observer. In this context, the unprimed quantities $g_1, g_2, g_3$ should not be confused with their primed relatives $g'_1, g'_2, g'_3$. The former, as explained below, constitute a system of coordinates for color space, and for each single observer, they are not stochastic variables.

For $\lambda_{\mathrm{t}}$ approximately between $430$ and $550$, nm no gains $(g'_1, g'_2, g'_3)$ achieve a perceptual match. If, however, the red primary is added to the target spectrum with a specific intensity $g'_3$, observers are able to find positive gains $g'_1$ and $g'_2$ to achieve the match. By convention, then, the CMF evaluated at one such target wavelength $\lambda_{\mathrm{t}}$ are defined by the trial average of the gains $(g'_1, g'_2, -g'_3)$. The negative sign of the last component indicates that the beam of wavelength $\lambda_3$ was added to the target field (as opposed to the matched field) with gain $g'_3$. 

Here we work under the hypothesis that observers select the gains so that the $\bm{k}'$ values produced by the mixture of primaries in the tested trial coincide with those of the target beam. The coincidence is expected to hold up to the expected standard deviation of both signals in multiple trials. In particular, the trial average of both $\bm{k}'$ vectors should coincide. Since the mean value of the Poisson distribution of Eq.~\ref{Pr} is $\alpha_i$, the $\bm{\alpha}$-values obtained by inserting Eq.~\ref{e4} into Eq.~\ref{e1} must coincide with those obtained when inserting the trial average of Eq.~\ref{e5} into Eq.~\ref{e1}. As a consequence, the trial-averaged gains $(g_1, g_2, g_3) = (\langle g'_1 \rangle, \langle g'_2 \rangle, \langle g'_3 \rangle)$ must be chosen so that \citep{Brainard2010a}
\begin{equation}
\sum_{j \in \{1,2,3\}} g_j \ I_j \ q_i(\lambda_j) = I_{\mathrm{t}} q_i(\lambda_{\mathrm{t}}), \ \ \ \ \ \ \forall i \in \{\mathrm{s}, \mathrm{m}, \ell\}.
\label{igualdad}
\end{equation}
The parameters $\beta_i$ describing the composition of the retina of the observer (Eq.~\ref{e1}) are cancelled out, so they do not appear in Eq.~\ref{igualdad}. As a consequence, the trial-averaged gains $g_j$ chosen by different observers do not depend on the retinal composition. Individual differences in the values of $g_i$, however, can still be expected, since some population variability in the shapes of the curves $q_i(\lambda)$ may remain.

The linear relation of Eq.~\ref{igualdad} between the column vector $\bm{g} = (g_1, g_2, g_3)^t$ of the gains and the column vector $\bm{t} = (q_{\mathrm{s}}(\lambda_{\mathrm{t}}), q_{\mathrm{m}}(\lambda_{\mathrm{t}}), q_\ell(\lambda_{\mathrm{t}}))^t$ of the target stimulus can be shortened by defining the matrix $Q$ with entries
\begin{equation} \label{e:matq}
Q_{ij} = q_i(\lambda_j), \ \ \ \ \mathrm{with} \ \ i \ \in \ \{\mathrm{s}, \mathrm{m}, \ell\} \ \ \mathrm {and} \ \ j \ \in \ \{1, 2, 3\}
\end{equation}
and the diagonal matrix $D$ with entries
\[
D_{jj} = \frac{I_j}{I_{{\mathrm t}}}.
\]
If the absorption probabilities $q(\lambda)$ vary from subject to subject, the matrix $Q$ can depend on the observer. The change of base matrix 
\begin{equation} \label{e:matc}
    C = Q \cdot D
\end{equation}
then relates $\bm{g}$ and $\bm{t}$: 
\[
C \ \bm{g} = \bm{t}.
\]
This equation can be solved uniquely for $\bm{g}$ for all non-singular $C$ matrices, yielding
\begin{equation} \label{e:efes}
\bm{g} = C^{-1} \ \bm{t}.
\end{equation}
The requirement of a non-singular $C$ is met by all triplets of non-coinciding primaries, as long as $D$ is invertible, that is, none of the beams is turned off. If two primaries, however, are close to each other, the matrix $C$ is close to singular, and one of its eigenvalues is close to zero. Unrealistically large gains $g_j$ may then be required. If less than three primaries are used, then $C$ is a rectangular matrix with more rows than columns, and cannot be inverted, implying that no match can be found. If, instead, more than three primaries are employed, $C$ is a rectangular matrix with more columns than rows, and the system has infinite solutions. One of the primaries can be obtained by combination of the remaining three, so whichever gain is assigned to that primary, could also have been distributed among the other three.

For each observer, Eq.~\ref{e:efes} defines a system of coordinates $(g_1, g_2, g_3)$ in which color can be represented. Since the vector $\bm{t}$ depends on the wavelength $\lambda_{\mathrm{t}}$ of the target beam, Eq.~\ref{e:efes} relates the CMFs $g_j(\lambda_{{\mathrm t}})$ to the spectral selectivity of the photon absorption process (through $Q$), and the properties of the three chosen primaries (through the wavelengths $(\lambda_1, \lambda_2, \lambda_3)$ and the associated intensities $(I_1, I_2, I_3)$ appearing in $D$. The three CMFs are linear combinations of the cone fundamentals $q_i(\lambda_{{\mathrm t}})$, and the coefficients of the linear combination, which depend on the three chosen primaries, define the change-of-base matrix $C$. 

If the population variability of the cone fundamentals $q_i(\lambda)$ is small, it makes sense to compare the population average of the gains $\bm{g}$ chosen by different observes with the prediction of Eq.~\ref{e:efes}. Figure~\ref{CMF}A 
displays the original CMFs reported by the CIE 1931, with the prediction for $g_i(\lambda_{\mathrm{t}})$ of Eq.~\ref{e:efes}, with $i \ \in \{1, 2, 3\}$. The diagonal elements  $I_i/I_{{\mathrm t}}$ of matrix $D$ were set to unity, which yields CMFs of unit Euclidean norm. The small discrepancy between the points and the curves could be due to the fact that the Stockman-Sharpe fundamentals are based on a different empirical set of color matching functions than those of the CIE 1931.

If instead of combining three monochromatic primaries, matching experiments are performed with light beams of arbitrary spectra $e_1(\lambda), e_2(\lambda), e_3(\lambda)$, the gains $g_j(\lambda_{{\mathrm t}})$ are still given by Eq.~\ref{e:efes}, but with a matrix $Q$ with elements $Q_{ij} = \langle q_i, e_j \rangle$, where the brackets represent a scalar product, defined as the integral in wavelength of the two involved functions. The resulting CMFs $g_j(\lambda_{{\mathrm t}})$ can still be obtained, and they still represent the gains of the three beams.

So far, the target beam was assumed to be monochromatic. If this restriction is relaxed, the spectrum $\tilde{I}_{{\mathrm t}}(\lambda)$ can be an arbitrary (non-negative) function. The linearity of Grassman's laws implies that the three trial averaged gains $(\tilde{g}_1, \tilde{g}_2, \tilde{g}_3)$ required to achieve the match are linear combinations of the CMFs obtained for monochromatic targets, that is, 
\begin{equation} \label{e:base}
\tilde{g}_j = \int g_j(\lambda) \ \tilde{I}_{{\mathrm t}}(\lambda) \ {\mathrm d}\lambda,
\end{equation}
where $g_j(\lambda)$ are given by Eq.~\ref{e:efes}. The values $\tilde{g}_1, \tilde{g}_2, \tilde{g}_3$ are the {\sl tri-stimulus values} of the beam $\tilde{I}_{{\mathrm t}}$, and constitute one possible system of coordinates in which the chromaticity of $\tilde{I}_{{\mathrm t}}(\lambda)$ is represented. Different choices of primaries result in different coordinate systems, since they yield different CMFs. 


\section{Primary colors producing convenient CMFs}
\label{s:triplets}

\begin{figure}[ht]
\begin{center}
\includegraphics[width=420pt]{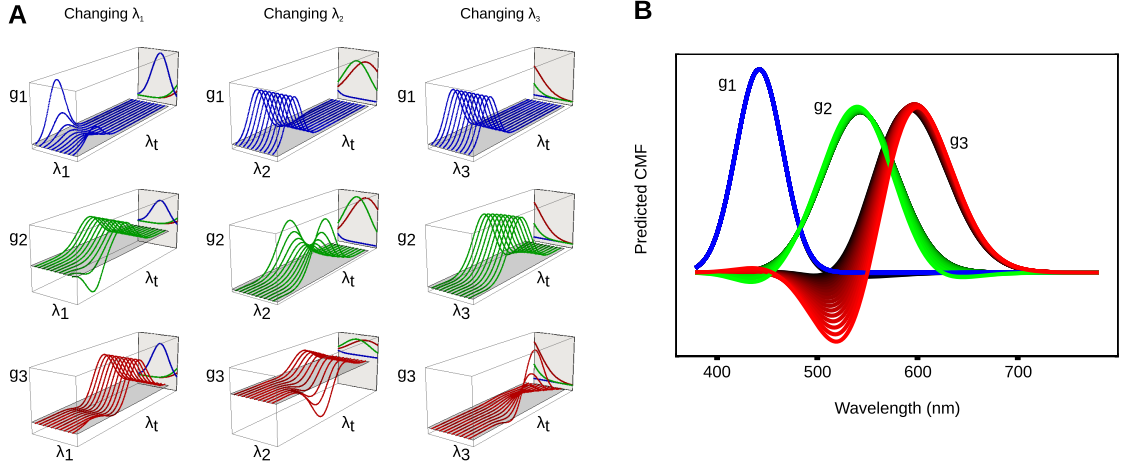}
\caption{Effect of varying the wavelengths $\lambda_1, \lambda_2, \lambda_3$ of the primary colors on the CMFs. A: In the first column, $\lambda_1$ varies in the interval $[390-480]$ nm, while $\lambda_2$ and $\lambda_3$ remain fixed at $526$ and $645$ nm, respectively. In the second column, $\lambda_2$ varies in $[490, 590]$ nm, while $\lambda_1$ and $\lambda_3$ remain fixed at $444$ and $645$ nm, respectively. In the third column, $\lambda_3$ varies in $[590, 690]$ nm, while $\lambda_1$ and $\lambda_2$ remain fixed at $444$ and $526$ nm, respectively. CMFs are displayed with $\lambda_t$ varying from $380$ to $779$ nm. Gray plane indicates $g_i = 0$. At the far end of each graph, the cone fundamentals are displayed in the same range of wavelengths as that of the primary that is varied. B: CMFs corresponding to triplets of primaries that interpolate linearly the triplet maximizing orthogonality ($\lambda_1 = 455$ nm, $\lambda_2 = 550$ nm, $\lambda_3 = 625$ nm, saturated colors) and the one maximizing positivity ($\lambda_1 = 380$ nm, $\lambda_2 = 510$ nm, $\lambda_3 = 775$ nm, black curves). } \label{f:primarios}.
\end{center}
\end{figure}
We now explore the dependence of CMFs on the selected primaries, and derive two triplets of primaries that produce CMFs that are particularly convenient. In Fig.~\ref{f:primarios}A, the dependence of the CMFs on the wavelengths $\lambda_1, \lambda_2, \lambda_3$ of the primaries is displayed. Two observations are relevant for the search of optimal primaries. First, as a given primary $\lambda_i$ moves away from the maximum of all cone fundamentals, the corresponding gain $g_i$ grows for all $\lambda_t$ values, since the observer must increase the intensity of the source to compensate for the decreased absorption probability. This effect is most evident in the 3-dimensional plots along the diagonal of the matrix of graphs in Fig.~\ref{f:primarios}A. The search for optimal primaries, hence, must be constrained to an interval within the visible range, the limits of which are set by the maximal gains compatible with ocular safety criteria.  Second, when two of the primaries are close enough to produce very similar activations of the cones (absorption probabilities displayed at the far end of each graph), one of the corresponding gains become negative. 

When the choice of primaries is only meant to produce CMFs that define a coordinate system (that is, whenever the actual execution of the color matching experiment is not required) unattainable primaries, often termed {\sl imaginary} primaries, may be employed. Imaginary primaries correspond to power spectra that contain negative values, and therefore, cannot be instantiated in reality. Such is the case, for example, of the primaries that underlie the $LMS$, the $RGB$ and the $XYZ$ coordinate systems. Equation~\ref{e:base} implies that the tri-stimulus values are the projection of the target spectrum $\tilde{I}_{{\mathrm t}}$ on the CMFs. Within this framework, the CMFs act as a base of the subspace of spectra that trichromats perceive.

\subsection{Maximizing the orthogonality of CMFs}

The choice of the first triplet is guided by the requirement of obtaining CMFs that be as orthogonal as possible. Coordinate systems constructed with orthogonal bases are desirable. If the base is orthogonal, whenever a set of stimuli exhibits correlations in their tri-stimulus values, those correlations can only be attributable to similarities in the spectra of the stimuli---and not to similarities in the primaries. Non-orthogonal bases, instead, produce correlations that also reflect similarities in the primaries.

The color matching functions reported by the CIE 1931 were not far from orthogonal. The scalar products of the normalized version of those curves were $\langle g_1, g_2 \rangle = 0.018$, $\langle g_2, g_3 \rangle = 0.044$, and $\langle g_3, g_1 \rangle = -0.015$, where the sub-indices $1, 2, 3$ refer to the primaries with wavelengths $435.8, 546.1$ and $700$ nm, respectively. The scalar products are small, but they can still be improved by diminishing $\langle g_2, g_3 \rangle$. 

The search for primaries that produce orthogonal CMFs has been undertaken before \citep{Thornton1999, Brill2007, Worthey2012}, by finding a linear transformation of some set of previously reported CMFs. However, the resulting primaries were imaginary. To produce an (almost) orthogonal base that corresponds to a realizable color-matching experiment, we performed an exhaustive numerical search of all triplets of monochromatic primaries between $380$ and $775$ nm, in steps of $5$ nm, calculated their CMFs through Eq.~\ref{e:efes}, and retained the triplet that minimized the function $\langle g_1, g_2\rangle^2 + \langle g_2, g_3\rangle^2 + \langle g_3, g_1\rangle^2$. The optimal triplet was $\lambda_1 = 455$ nm, $\lambda_2 = 550$ nm, and $\lambda_3 = 625$ nm. The main difference with the CIE 1931 primaries is that the wavelength of the red beam is diminished.

The resulting CMFs are displayed in Fig.~\ref{CMF}B, and the most noticeable difference with Fig.~\ref{CMF}A, is that  $g_2$ contains larger negative regions flanking both sides of its maximum, thereby diminishing the overlap with $g_1$. The inner products between the resulting CMFs are $\langle g_1, g_2 \rangle = 0.012, \langle g_2, g_3 \rangle = 0.011, \langle g_3, g_1 \rangle = -0.01$. 

\subsection{Maximizing positivity of CMFs}

We also searched for primaries that produce CMFs with maximal domain of positive values. Such primaries are the optimal choice when attempting to construct metamers of monochromatic beams with the largest possible range of target wavelengths, since the negative portion of CMFs reflect a failure to construct the target percept. This request is relevant, for example, when choosing the LEDs of computer screens. Again, we performed a numerical, exhaustive search of monochromatic primaries, and maximized the sum of the domains where the resulting CMFs were positive. The optimal triplet had wavelengths $\lambda_1 = 380$ nm, $\lambda_2 = 510$ nm, and $\lambda_3 = 775$ nm. In this case, the wavelengths are more separated from one another than in the original CIE 1931 primaries, and reached the minimal and maximal values allowed by the search. Clearly, if no restriction is imposed on the amplitude of the gains, even more separated primaries would produce still more positive CMFs, so as to minimize coactivation of cones of different types. The optimal normalized CMFs are displayed in Fig.~\ref{CMF}C. Negative values could not be avoided for $g_1$ and $g_3$, but the reached values were small ($-0.003$ and $-0.002$, respectively), so it may be hypothesized that for those wavelengths, replacing a negative gain by zero would produce a minimal perceptual shift.


\subsection{Tradeoff between orthogonality and positivity}

Above we obtained two sets of primaries that produced CMFs that were either maximally orthogonal, or maximally positive. One may then wonder whether these two requisites may be attained simultaneously. The CMFs are linear combinations of the cone fundamentals (see Eq.~\ref{e:efes}), and as the $q_\mathrm{m}(\lambda)$ and $q_\ell(\lambda)$ are largely overlapping, no triplet of primaries can produce CMFs that occupy non-overlapping intervals of wavelengths. We must therefore accept that CMFs inevitably overlap. Hence, the only way to obtain orthogonal or nearly orthogonal CMFs is to allow for negative values, implying that positivity and orthogonality rival a tug of war that cannot be won by both: if positivity increases, orthogonality diminishes, and vice versa. Therefore, in a given problem, the optimal triplet should be chosen depending on the cost/benefit ratio of either positivity or orthogonality in the application at hand. In Fig.~\ref{f:primarios}B, we show the CMFs resulting from a collection of triplets of primaries that vary linearly between the optimal for positivity and the optimal for orthogonality. As we move from the first to the second, the only CMF that changes conspicuously is the one corresponding to the primary with longest wavelength. 


\section{Trial-to-trial variability and correlations of CMF}
\label{s:varcovar}

We now turn to a different goal, namely, to provide a principled derivation of the trial-to-trial variability and correlation structure of the single-subject CMFs, capturing the dispersion and the structure of the observer's responses. In our derivation, the source of variability is the stochastic nature of photon absorption (Eq.~\ref{Pr}). We are aware of the existence of additional sources of variability. Still, here the aim is to assess how much of the experimental variability can be accounted for, taking only the stochasticity of photon absorption of Eq.~\ref{Pr} into account. The advantage of describing photon absorption alone, is that the associated probability distribution (Eq.~\ref{Pr}) can be derived from first principles \citep{Fonseca2016}.


\subsection{Deriving variances and covariances}
\label{derivation}

\begin{figure}[ht]
\begin{center}
\includegraphics[scale = 0.8]{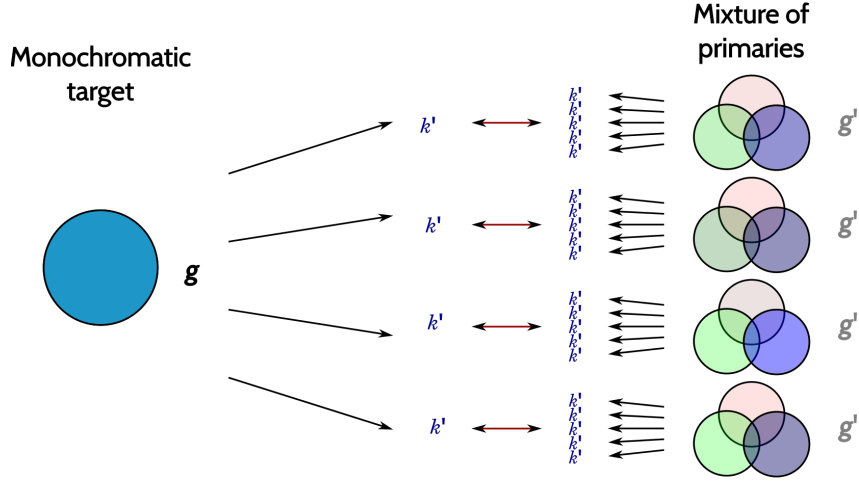}
\caption{A stimulus of a given $\bm{g}$ vector produces activations $\bm{k}'$ that vary from trial to trial. Observers choose a triplet of gains $\bm{g}'$ so that the mixture of primaries produces a signal $\bm{k}'$ that is similar to that produced by the monochromatic target beam. Red arrows represent the comparison. We work under the hypothesis that the collection of $\bm{k}'$ vectors generated by each of the two stimuli in multiple trials of the matching condition define two probability distributions $P(\bm{k}'|\lambda_{\mathrm{t}}) = P(\bm{k}'|\text{mixture})$. The equality of the two distributions implies the equality of their means. Therefore, the coordinates $(g_1, g_2, g_3)$ of the monochromatic target obtained from Eq.~\ref{e:efes} must coincide with the trial average of the selected $(g'_1, g'_2, g'_3)$. In single trials, a match is enacted when the $\bm{k}'$ vectors elicited by the two stimuli differ in no more than the typical trial-to-trial variability of the signals of a single stimulus, here represented by the black arrows. Therefore, different trials with a single monochromatic target produce different $\bm{g}'$ vectors (four of them displayed in the figure). 
\label{f:esquema}}
\end{center}
\end{figure}

The matching experiment involves the comparison of the neural responses produced by two stimuli: the target stimulus and the mixture of primaries. Both  excite cones, and produce noisy signals in the photoreceptor layer that propagate downstream. We assume that subjects report a match whenever the discrepancy of the two signals at some level of processing involved in decision making is at most of the order of the trial-to-trial variability expected in each single signal. In other words, we work under the hypothesis that the inherent noise with which the system responds to stimuli in the tested region of color space determines the precision demanded on the comparison between the neural activities of the two beams for the match to be accepted as such. Figure~\ref{f:esquema} illustrates the logical scheme.

The matching procedure can be conceived as a noisy measurement of the target stimulus, in a causal chain 
\[ \begin{array}{c} \text{Target stimulus} \\ (g_1, g_2, g_3)\end{array} \to \begin{array}{c}\text{Cone activity}  \\ \text{produced by target} \end{array} \to \text{Further processing} \to (g'_1, g'_2, g'_3).
\]
In this chain, the second stage is the cone activity produced by the target stimulus, whereas the cone activity produced by the mixture of primaries is hidden in ``further processing'', as an element of the matching strategy. This asymmetry between the two beams is only apparent, since exactly the same chain can be constructed placing the mixture of primaries on the left side, and the target stimulus in ``further processing''. We here focus on the target stimulus because in all the classical experiments discussed here, CMFs are reported as a function of the target wavelength. Moreover, when multiple trials are considered, the wavelength is kept fixed whereas the gains are tuned. A different experiment with fixed gains and an adjustable wavelength is of course also conceivable. In the classical papers, the gains $(g'_1, g'_2, g'_3)$ constitute the output of the chain, and instantiate one possible measuring apparatus representing the target wavelength. The comparison of the true $(g_1, g_2, g_3)$ obtained from Eq.~\ref{e:efes} with the single-trial estimated $(g'_1, g'_2, g'_3)$ defines an estimation error. The ideal observer paradigm employed in the rest of the paper is a theoretical construct that bounds the minimal mean quadratic error that any estimator of the target stimulus based on the cone activity must have. The mean quadratic error of the matching experiment, which instantiates one particular estimator among the many possible ones, cannot be less than the minimal. 

The trial-to-trial fluctuations of $\bm{g}'$ are captured by the $3 \ \times \ 3$ mean quadratic error matrix $E$ of entries
\[
E_{ab}(\bm{g})= \left\langle \left[g'_a(\bm{k}')- g_a \right] \  \left[g'_b(\bm{k}')- g_b \right] \right\rangle,
\]
where the brackets represent an expectation value weighted with $P(\bm{k}' | \bm{g})$, the sub-indexes $a$ and $b$ vary in the set $\{1, 2, 3\}$. The diagonal elements of $E_{jj}$ represent the variances of the measured $g'_i$ values, and the off-diagonal elements $E_{ab}$, the covariances. More precisely, the matrix $E(\bm{g})$ is the quadratic form of an ellipsoid that in each direction encompasses the central $68\%$ of the trial-to-trial fluctuations in the estimated $\bm{g}'$. The principal axes of the ellipsoid are given by the eigenvectors of $E(\bm{g})$, and their lengths, by the square root of the corresponding eigenvalues.

The Cr\'amer-Rao bound \citep{Rao1945, Cramer1946, Cover2012} states that the mean quadratic error $E$ of any unbiased estimator is bounded from below by the inverse of the Fisher Information $J(\bm{g})$, a $3 \ \times \ 3$ matrix of entries 
\begin{equation} \label{Jcmf}
J_{ab}(\bm{g}) = - \left\langle \frac{\partial^2 \ln P(\bm{k}' | \bm{g})}{\partial g_a \ \partial g_b} \right\rangle.
\end{equation}
The Fisher Information  matrix is the metric tensor with which infinitesimal distances in color space can be calculated, such that traversing a unit of distance in color space modifies the distribution of $\bm{k'}$ vectors in a fixed amount \citep{Amari2000}. Several studies have used the notion of Fisher Information to describe psychophysical experiments in chromatic perception \cite{Zhaoping2011, Fonseca2016, Fonseca2018, Fonseca2019}. The Cr\'amer-Rao bound reads
\begin{equation} \label{CR}
E \cdot J \ge \mathbb{1},
\end{equation}
and states that all the eigenvalues of the matrix product $E \cdot J$ must be larger or equal than unity. It implies that, inasmuch as $J$ is associated to the notion of information, $J^{-1}$ is associated to the notion of minimal mean quadratic estimation error. The larger the information, the smaller the error, and vice versa. The fact that Eq.~\ref{CR} is expressed in matrix form means that the bound is directional. In other words, the mean quadratic error may take different values along different directions: Along the eigenvectors of $E$, the error is equal to the corresponding eigenvalues. Equation~\ref{CR} is only valid for unbiased estimators, that is, those for which $\langle \bm{g}'(\bm{k}') \rangle = \bm{g}$. A more complex formula is required in the biased case \citep{Cover2012}. However, in our case, by definition, $\bm{g} = \langle\bm{g}'\rangle$.

Equation~\ref{CR} is an inequality, so the Fisher Information can be employed to bound, but not to calculate, the mean quadratic error. Even so, in this paper we assume that the equality holds, and derive the mean quadratic error analytically as 
\begin{equation} \label{e:igualdad}
E \approx J^{-1},
\end{equation}
since $J$ can be obtained analytically from Eqs.~\ref{Jcmf} and \ref{Pr}. The assumption is only valid if all subsequent processing stages, downstream from photon absorption, preserve the information encoded by the vector $\bm{k}'$. In 2016, we showed that the mean quadratic error obtained by assuming that the equality holds captures 87\% of the variance of behavioral discrimination experiments \citep{Fonseca2016}. Assuming the equality, hence, seems to be justified up to a reasonable degree. Continuing with this line of thought, we here explore the consequences of this assumption in the mean quadratic error of behavioral matching experiments.

The Fisher Information matrix was obtained in \cite{Fonseca2016}, in the $LMS$ coordinate system, obtaining a diagonal matrix of entries
\begin{equation} \label{e:jalfa}
J(\bm{\alpha})_{ab} = \frac{1}{\alpha_a} \ \delta_{ab},
\end{equation}
where $\delta_{ab}$ is the Kronecker delta symbol. To predict the trial-to-trial fluctuations in matching experiments, this tensor must be transformed to the $\bm{g}$ coordinate system. To that end, we define the diagonal matrix $B$ with entries
\[
B_{ij} = \beta_i \ \delta_{ij},
\]
containing the fractions $\beta_i$ of each type of cones. The coordinate transformation between $\bm{\alpha}$ and $\bm{g}$ is
\begin{equation} \label{e:transf}
 \bm{\alpha} = B \cdot C \ \bm{g}.
\end{equation}
Consequently, the transformation rule for the metric tensor is \citep{Fonseca2016}
\begin{equation} \label{e:jg}
J(\bm{g}) = \left(B \cdot C\right)^t \cdot J(\bm{\alpha}) \cdot \left(B \cdot C\right).
\end{equation}
Inserting Eq.~\ref{e:jalfa} into Eq.~\ref{e:jg}, using the expression \ref{e:transf}, and solving for $J(\bm{g})$, the Fisher Information can be obtained analytically,
\begin{equation} \label{e:jgexplicito}
J(\bm{g})_{ab} = I_a \ I_b \ \sum_{i \ \in \ \{\mathrm{s}, \mathrm{m}, \ell \}} \frac{\beta_i \ q_i(\lambda_a) \ q_i(\lambda_b)}{\sum_{j = 1}^3 q_i(\lambda_j) \ I_j \ g_j}.
\end{equation}
An important consequence of Eq.~\ref{e:jgexplicito} is that, whenever two of the primaries $\lambda_a$ and $\lambda_b$ are close enough such that at least one of the cone fundamentals $q_i(\lambda)$ is different from zero on both primaries, the Fisher Information contains non-vanishing off-diagonal elements. The cone fundamentals cover a fairly broad range of wavelengths. Moreover, the triplets of primaries usually employed often contain a green and a red light, the absorption probability of which does neither vanish for $M$ nor for $L$ cones, yielding a nonzero green-red matrix element of $J$. Therefore, neither $J$ nor $J^{-1}$ are diagonal matrices, giving rise to tristimulus values with correlated variabilities. Hence, independent photon absorption in $S$, $M$ and $L$ cones does not imply that $g_1, g_2$ and $g_3$ be independent from one another. The correlated action of the gains becomes manifest when the conditional probability distribution $P(\bm{k}'|\bm{\alpha})$ of Eq.~\ref{Pr} is re-written as a function of $g_1, g_2, g_3$, that is, 

\[
P(\bm{k}'|\bm{g}) = \prod_{i \in \{\mathrm{s}, \mathrm{m}, \ell\}} \exp\left[-\frac{\beta_i}{I_{\mathrm{t}}} \sum_{j = 1}^3 I_j \ q_i(\lambda_j) \ g_j \right] \  \frac{\left[-\frac{\beta_i}{I_{\mathrm{t}}} \sum_{j = 1}^3 I_j \ q_i(\lambda_j) \ g_j \right]^{k'_i}}{k'_i!}.
\]
This expression cannot be factored out into a product of three functions, each containing a single $g_i$.

\begin{figure}[ht]
\begin{center}
\includegraphics[width=420pt]{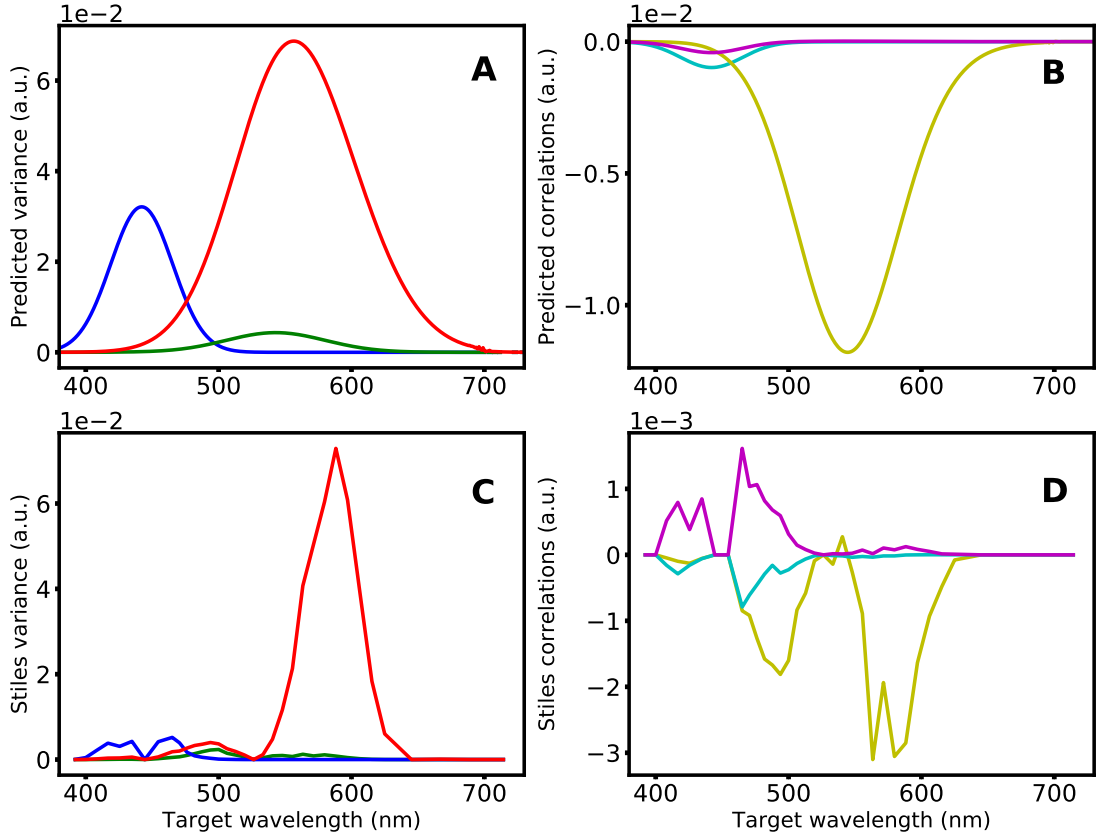}
\caption{A: Normalized theoretical prediction for the variance of the Stiles CMFs \citep{Stiles1959} obtained with primaries $\lambda_1=444$ nm, $\lambda_2=526$ nm, $\lambda_3 = 645$ nm, as a function of the target wavelength $\lambda_{{\rm t}}$. The three curves are the diagonal terms of the inverse of the matrix in Eq.~\ref{e:jfinal}. Color code as in Fig.~\ref{CMF}A. B: Correlations between the CMFs, obtained from the off-diagonal terms of the inverse of the matrix in Eq.~\ref{e:jfinal}. Cyan: correlation between $g_1$ and $g_2$. Yellow: correlation between $g_2$ and $g_3$. Magenta: correlation between $g_3$ and $g_1$. D: Variances and correlations obtained from multiple subjects performing the color-matching experiment \citep{Stiles1959}. Same color convention as in A and B.} \label{corr}
\end{center}
\end{figure}
If the color-matching experiment is performed with monochromatic target stimuli (Eq.~\ref{e4}), the Fisher Information matrix of Eq.~\ref{e:jgexplicito} reduces to
\begin{equation} \label{e:jfinal}
J(\bm{g})_{ab} = \frac{I_a \ I_b}{I_{\mathrm{t}}} \ \sum_{i \ \in \ \{\mathrm{s}, \mathrm{m}, \ell \}} \frac{\beta_i \ q_i(\lambda_a) \ q_i(\lambda_b)}{q_i(\lambda_{\mathrm{t}})}.
\end{equation}
Writing the intensities $(I_1, I_2, I_3)$ in units of $I_{\mathrm{t}}$ reveals that the Fisher Information is linear in the target intensity $I_{\mathrm{t}}$. Therefore, the variance of the responded $g'_j$ is inversely proportional to the total light intensity employed in the experiment.

The Fisher Information matrix of Eq.~\ref{e:jfinal} can be inverted to yield the mean quadratic error under the assumption of Eq.~\ref{e:igualdad}. In Fig.~\ref{corr}A, the diagonal elements $E_{aa}$ (the variances) are displayed for the primaries employed by \citet{Stiles1959}, and a retinal composition of $\beta_{\mathrm{s}} = 0.05, \beta_{\mathrm{m}} = 0.45, \beta\ell = 0.5$, which is quite typical for human trichromats.

The ratios $I_j / I_{\mathrm{t}}$ were set to $1$, in order for the resulting CMFs to have unit norm. The global scaling factor $I_{\mathrm{t}}$ was set to $0.014$, in order for the maximum of the variance in $g'_3$ (peak of the red curve in Fig.~\ref{corr}A) to equate the experimental height (Fig.~\ref{corr}C, see below).

The off-diagonal elements $E_{ab}$ can be seen in Fig.~\ref{corr}B, showing that all correlations are negative, and they tend to be particularly significant in those regions of the spectrum where the two corresponding CMFs overlap. Negative correlations imply that if, in one particular trial, the observer sets one of the gains above average, they are likely to set the other two below average, at least, if there is an overlap between the corresponding CMFs.   

Correlations are maximal (in absolute value) for the green and red primaries, when $\lambda_{\mathrm{t}}\approx 550$ nm. If $\beta_{\mathrm{m}} \approx \beta_\ell$, and $I_2 \approx I_3$, at this target wavelength the pronounced similarity between $q_{\mathrm{m}}$ and $q_{\ell}$ implies that in the base $(g_1, g_2, g_3)$ the Fisher information takes the form
\[
\left.J(\bm{g})\right|_{\lambda_{\mathrm{t}} \ \mathrm{near} \ 550 \ \mathrm{nm}} \approx  \left(\begin{array}{ccc} \mu & 0 & 0 \\ 0 & \nu & \rho \\ 0 & \rho & \nu \end{array} \right),
\]
with $\mu, \nu$ and $\rho$ defined in Eq.~\ref{e:jfinal}. This matrix is diagonalized by the eigenvectors $(1, 0, 0), (0, 1, 1)$ and $(0, 1, -1)$. Therefore, the combinations $g_2 + g_3$ and $g_2 - g_3$ are uncorrelated coordinates.  It should be noted that the Fisher information derived from photon-absorption is smaller along the direction $g_2 - g_3$, thereby predicting higher variance, and therefore, poorer discriminability. This result contradicts experimental findings that discriminability is lower along the direction $g_2 + g_3$ \citep{Wandell1985, Wandell1995}. Downstream processing stages are probably responsible for this effect, but they are not included in the present model.


\subsection{Variances and covariances for different retinas}
\label{s:retinas}

\begin{figure}
\begin{center}
\includegraphics[width=420pt]{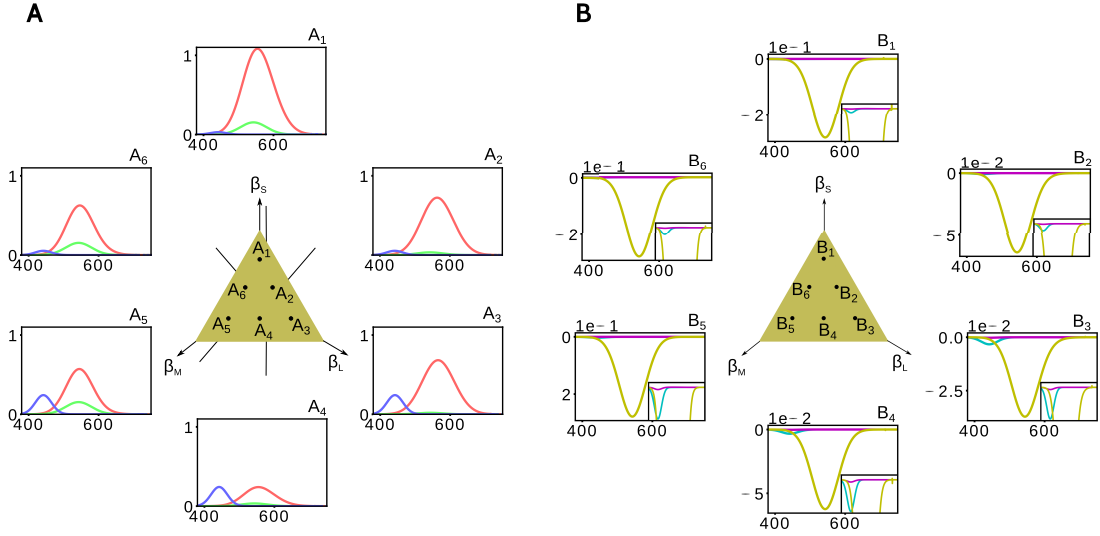}
\caption{Effect of varying the retinal composition on the (A) variances, defined as the diagonal elements of $J^{-1}$, and (B) the covariances, defined as the off-diagonal elements of $J^{-1}$ of the CMFs. Since $\beta_{\mathrm{s}} + \beta_{\mathrm{m}} + \beta_\ell = 1$, the vector $\bm{\beta}$ belongs to the plane normal to the direction $(1, 1, 1)$. All three $\beta_i$ are non-negative, so as the retinal composition varies, the vector $\bm{\beta}$ moves on the triangular, planar region depicted at the center of each figure. Panels 1-6 display the variances and covariances of each corresponding $\bm{\beta}$-value marked in the central triangle, such that $\bm{\beta}^t = (\beta_{\mathrm{s}}, \beta_{\mathrm{m}}, \beta_\ell) = (0.8, 0.1, 0.1)$ at the top, and $(0.45, 0.1, 0.45), (0.1, 0.1, 0.8), (0.1, 0.45, 0.45), (0.1, 0.8, 0.1)$ and $(0.45, 0.45, 0.1)$, respectively, as we rotate clockwise. In (A), blue, green and red curves correspond to $J_{11}^{-1}, J_{22}^{-1}$ and $J_{33}^{-1}$, respectively. In (B), cyan, yellow and magenta curves correspond to $J_{12}^{-1}, J_{23}^{-1}$ and $J_{31}^{-1}$, respectively. Insets enlarge the figures so that the cyan and the magenta covariances are also visible. All curves are depicted as a function of the target wavelength $\lambda_{\mathrm{t}}$. Primary colors $\lambda_1, \lambda_2, \lambda_3$ and beam intensities $I_1, I_2, I_3$ as in Fig.~\ref{corr} \label{f:betas}.}
\end{center}
\end{figure}

The Fisher Information matrix bears an explicit dependence on $(\beta_{\mathrm{s}}, \beta_{\mathrm{m}}, \beta_\ell)$, implying that observers with different retinal composition respond with trial-to-trial fluctuations of varying structure. Fig.~\ref{f:betas} contains the variances and covariances of all possible retinal compositions. However, the population variability of the proportion $\beta_{\mathrm{s}}$ of $S$ cones fluctuates within a narrow range, between 1\% and 5\% \citep{Wyszecki1980,Roorda1999,Sabesan2015}. The relative fraction $\beta_\ell/\beta_{\mathrm{m}}$ of $M$ to $L$ cones is highly variable, ranging between 0.3 and 10 \cite{Kremers2000,Carroll2002,He2020}. Therefore, the range of $\bm{\beta}$ values that are close to realistic correspond to the lower panels (3, 4, 5) of  Fig.~\ref{f:betas}. In this range, the ratio $\beta_{\ell}/\beta_{\mathrm{m}}$ modulates the magnitude of the variance of $g'_3$ and the covariance between $g'_2$ and $g'_3$. In the calculations, it must be born in mind that as the vector $\bm{\beta}$ reaches the boundary of the triangular region, the matrix $J$ becomes close to singular, so its inversion may yield numerical problems.

A few studies \citep{Dartnall1983,Burns1993} have also described a subject-to-subject variability of the peak wavelength of the cone fundamentals $q_i(\lambda)$,  
\begin{figure}
\begin{center}
\includegraphics[width=420pt]{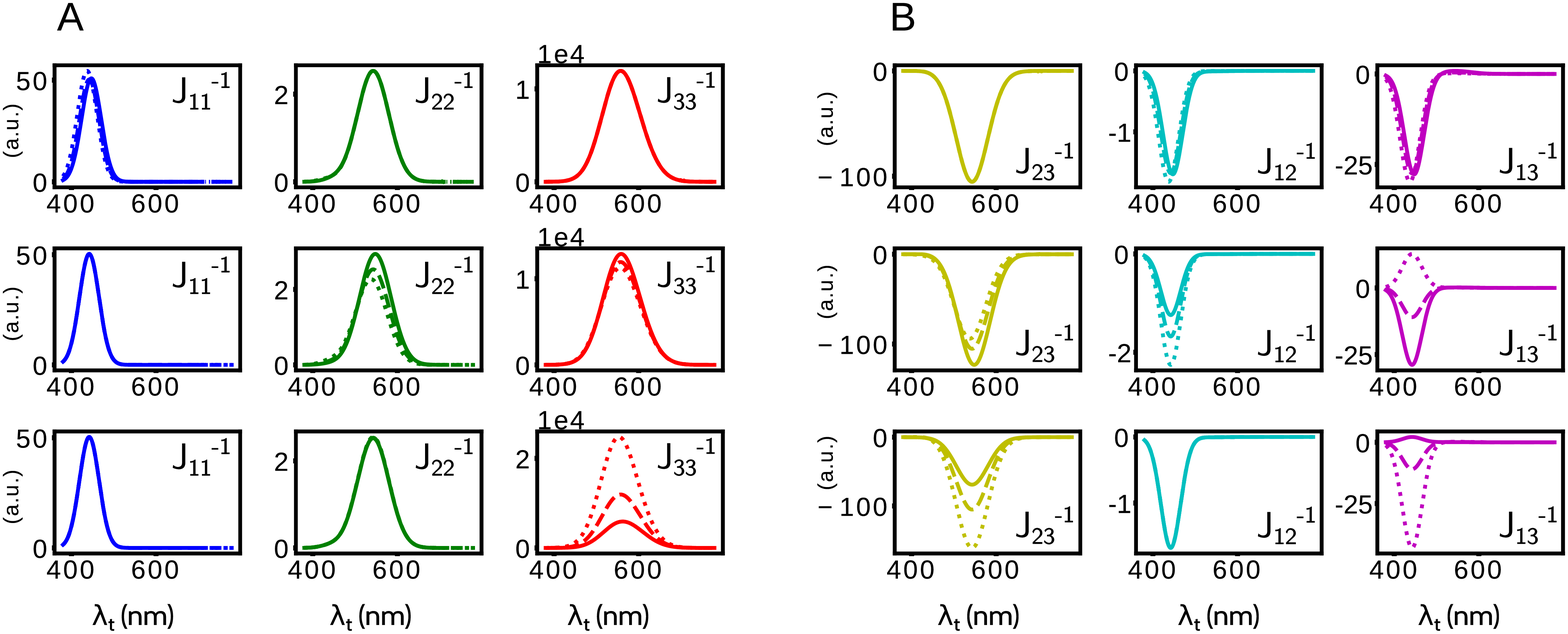}
\caption{Effect of varying the peak wavelength of the cone fundamentals $q_i(\lambda)$ on the (A) variances, and (B)  covariances of the CMFs, as measured by the (A) diagonal and the (B) off-diagonal terms of the matrix $J^{-1}$, for different target wavelengths $\lambda_{\mathrm{t}}$. Each $q_i(\lambda)$ was modeled as $\exp[-(\lambda - \lambda_i^{\mathrm{max}})^2/\sigma_i^2]$, with $\lambda_{\mathrm{s}} = 442.1 \pm 5$ nm, $\lambda_{\mathrm{m}} = 542.8 \pm 5$ nm, $\lambda_{\ell} = 442.1 \pm 5$ nm, $\sigma_{\mathrm{s}} = 32.96$ nm, $\sigma_{\mathrm{m}} = 52.8$ nm and $\sigma_\ell = 64.76$ nm. Top, middle and bottom rows display the results of shifting $\lambda_i^{\mathrm{max}}$ for $i = \mathrm{s}, \mathrm{m}$ and $\ell$ cones, respectively (only a single cone fundamental is shifted in each row). Dotted, dashed and full lines indicate $\lambda_i^{\mathrm{max}} - 5$ nm, $\lambda_i^{\mathrm{max}}$ and $\lambda_i^{\mathrm{max}} + 5$ nm, respectively . Color code to represent the variance of each CMF (left) or the covariance of different pairs of CMFs same as in Fig.~\ref{corr}.  Primary colors $\lambda_1, \lambda_2, \lambda_3$ and beam intensities $I_1, I_2, I_3$ as in Fig.~\ref{corr} \label{f:cf}.}
\end{center}
\end{figure}
reporting shifts of up to 5 nm. These shifts produce a negligible effect on the shape of the CMFs themselves (not shown), but a non-negligible effect on the variances and covariances. Shifting $\lambda_i^{\mathrm{max}}$ affects the magnitude of the variance of $g'_i(\lambda_{\mathrm{t}})$ (Fig.~\ref{f:cf}A), but has virtually no effect on the variance of the other two CMFs (only the panels on the diagonal display visible differences). The covariances are more sensitive to the peak wavelengths. Shifting $\lambda_{s}^{\mathrm{max}}$ affects the covariances between $g'_1$ and $g'_2$, as well as that between $g'_1$ and $g'_3$, though the effect is minor (top panel of Fig.~\ref{f:cf}B). Shifting $\lambda_{m}^{\mathrm{max}}$ affects all three covariances, even the one between $g'_1$ and $g'_3$ (middle panel of Fig.~\ref{f:cf}A).  Shifting $\lambda_\ell^{\mathrm{max}}$ affects the covariance between $g'_1$ and $g'_3$, and that between $g'_2$ and $g'_3$, but not that between $g'_1$ and $g'_2$, as expected. Quite remarkably, the covariance between $g'_1$ and $g'_3$ sometimes even inverts its sign, from negative to positive. The inversion befalls when the shift in $\lambda_i^{\mathrm{max}}$ is such as to diminish the (inescapably strong) correlation between $g'_2$ and $g'_3$. Apparently, a small release in the almost enslaving relation between $g'_2$ and $g'_3$ suffices for the negative correlation between $g'_2$ and the other two signals to produce, by transitivity, a positive correlation between $g'_1$ and $g'_3$.


\subsection{Primary colors yielding high information}
\label{s:primariosinformativos}

In Sect.~\ref{s:triplets}, we searched for triplets of colors that produced CMFs $g_j(\lambda)$ with certain convenient properties: orthogonality or positivity. These advantages referred to the trial-averaged CMFs. We now derive the optimal triplet in terms of maximizing the Fisher information which, in view of the Cr\'amer-Rao bound, imposes constraints on the mean quadratic fluctuations of the responses.

Equation~\ref{Jcmf} implies that the components of the Fisher information are large whenever a small change in a gain produces a detectable change in the conditional probability distribution $P(\bm{k}'|\bm{g})$. A change is considered to be detectable if it is larger than the inherent variability of absorption Poisson noise. The tensorial nature of $J$ assesses the degree up to which the three gains $(g'_1, g'_2, g'_3)$ are able to control three independent directions of the $\bm{k}'$. The three eigenvalues of the Fisher information evaluate this effectiveness in the three orthogonal directions of the corresponding eigenvectors. To employ a measure that comprises all three directions, we here choose to quantify the effectiveness of a triplet of primaries $(\lambda_1, \lambda_2, \lambda_3)$ with the absolute value of the determinant of the Fisher information, defined as the product of the three eigenvalues. The Crámer-Rao bound states that this determinant is an upper limit of the inverse of the volume occupied by the cloud of $\bm{g}'$ points obtained in repeated trials by a single subject.

Equation~\ref{e:jg} implies that
\[
\det J(\bm{g}) = \det \left[BQD\right]^2 \ \det J(\bm{\alpha}). 
\]
The primary colors $\lambda_1, \lambda_2, \lambda_3$ only enter into this determinant through the matrix $Q$ defined in Eq.~\ref{e:matq}. Since this matrix does not depend on the tested position of color space (set by $\bm{\alpha}$, or equivalently, $\bm{g}$), the optimal triplet of primaries is the same for the whole space -- even though $J$ varies throughout color space.

The determinant $|\det(Q)|$ is maximal when (a) the columns of $Q$ are maximally orthogonal to each other, and (b) they have maximal length. These two conditions ensure that the transformation $\bm{\alpha} \to \bm{g}$ maximizes the ratio of differential volumes. The first requirement favors primary colors that are maximally separated from each other, so that each primary excite a single cone fundamental. However, too separate primaries are incompatible with the second requirement, since the absorption probabilities $q_{\mathrm{s}}(\lambda)$ and $q_\ell(\lambda)$ diminish rapidly for wavelengths below $400$ nm, or above $600$ nm. Therefore, a tradeoff between the two requirements arises.

Maximizing the orthogonality is tantamount to requiring that all three components of the $\bm{\alpha}$-vector (and thereby, the $\bm{k}'$-vector) be sensitive to the gains. Triplets of primaries that are too similar do not accomplish this feat. They achieve a precise control of one direction of the $\bm{\alpha}$-space (implying that one of the eigenvalues of $J(\bm{g})$ is large), but the other two remain unattended, resulting in two small eigenvalues. This means that there are many combinations of the gains that are mapped to essentially the same region of the $\bm{\alpha}$-space, and produce a negligible perceptual effect. In turn, maximizing the lengths of the columns of $Q$ implies that the primaries are sensitive knobs. When the components of the matrix $Q$ are large, from the equation $\bm{t} = Q D \bm{g}$, we deduce that a small change in the gains produces a large displacement in the $\bm{\alpha}$ space. For the elements $Q_{ij}$ to be large, the primary colors must be within the range of wavelengths in which al least one type of cone responds effectively. 

To find the optimal triplet, we made a numerical search, and concluded that the most reliable primaries were $\lambda_1 = 442.1$ nm (coinciding with the wavelength that maximizes the absorption probability of $S$-cones), $\lambda_2 = 535.2$ nm, and $\lambda_3 = 595.7$ nm. These last two wavelengths do not maximize the absorption probability of $M$ and $L$ cones, which peaks at $542.8$ nm and $568.2$ nm, respectively. Although these shifts imply matrix entries $Q_{ij}$ that are smaller in magnitude than would appear at the peak wavelengths, they are required to orthogonalize the second and third columns of $Q$, in view of the pronounced overlap of the cone fundamentals of $M$ and $L$ cones.

\subsection{Intra-observer vs. inter-observer variability}
\label{s:experiments}

Some previous studies have addressed the subject-to-subject variability of color-matching experiments \citep{Stiles1959,Wyszecki1971,Webster1988,Alfvin1997,Fairchild2013,Fairchild2016,Asano2016b,Asano2016a,Emery2017,Murdoch2019, Emery2019}, and only a few have explored the trial-to-trial variability of the responses of a single subject \citep{Wyszecki1971,Alfvin1997,Sarkar2010,Asano2015}. The two types of variability derive from different sources. Subject-to-subject variability is mainly due to individual differences in biophysical and physiological properties, and describes the degree of agreement in the percept produced by a given stimulus in a population of observers. Trial-to-trial variability, instead, reflects the inherent uncertainty with which a given observer perceives a given stimulus, and stems from noisy processes both outside and inside the visual system. 

The inter-subject variability, has been more exhaustively characterized, probably for commercial purposes, and can be depicted as a function of wavelength (Fig.~\ref{corr}C and D). Theoretical studies \citep{Fairchild2013,Asano2015,Asano2016b,Murdoch2019} on the inter-subject variability take into account individual differences in lens and macular pigment density, retinal composition (matrix $B$), and variations in the shape of the cone fundamentals $q_{\mathrm{s}}(\lambda), q_{\mathrm{m}}$ and $q_\ell(\lambda)$. The relevance of these parameters was first identified in the factor analysis performed by \citet{Webster1988} of the  CMFs collected by \citet{Stiles1959} on a population of 49 subjects \citep{Stiles1959}, and that are available online (Fig.~\ref{corr}C and D).

Unfortunately, we lack experimental data on the trial-to-trial fluctuations of a single observer, at least, beyond crude estimations performed with very few samples. The results of \citet{Wyszecki1971} are difficult to interpret, since the variability of a single subject in different sessions (separated by several weeks or months) is considerably larger than one obtained in a single session of multiple trials, suggesting that some experimental conditions may have changed from one session to the next. Moreover, the results differ significantly from those of MacAdam, suggesting that the recommendation to continuously sweep the gaze through the stimuli may have produced additional effects. Experiments estimating the intra-observer variability from $3$ matches performed by each subject were published in the PhD Dissertation of Yuta Asano \citet{Asano2015}. This trial-to-trial variability  was approximately half the inter-observer variability, in accordance with earlier estimations \citep{Alfvin1997,Sarkar2010}. Importantly, the variability was estimated for a discrete collection of non-monochromatic target colors, so it cannot be displayed as a function of the target wavelength. 

To our knowledge, the present study provides the first analytical derivation of the trial-to-trial variability of the CMFs, deduced from one well identified source of noise: The Poissonian nature of photon absorption. The theoretical framework is exactly the same as the one employed by \citet{Zhaoping2011} and \citet{Fonseca2016}, with the only difference that the calculation was here specifically adapted to a color matching experiment performed with three primaries of fixed wavelengths. The result can be displayed as a function of the target wavelength (Fig.\ref{corr}A and B). Since our results cannot be reliably compared with experimental data recorded with multiple trials in a single observer, we compare them with those obtained with a single trial of multiple observers, understanding that differences are expected, due to the diverse sources of variability. 

The experimental result of the population variances exhibit a marked peak for the red primary (Fig.~\ref{corr}C), the position of which only roughly coincides with the prediction of the theoretical variance for a single observer (Fig.~\ref{corr}A). Therefore, part of the variance reported in the experimental result could potentially stem from intra-subject variability. The experimental variances of the other two primaries (green and blue curves in Fig.~\ref{corr}A) are too noisy to be useful. However, the relative size of the blue curve (compared to the red) in the theoretical result does not coincide with the experimental relation. Therefore, the population variability present in Fig.~\ref{corr}C and absent from Fig.~\ref{corr}A probably affects differentially the two primaries. No conclusions can be drawn about the absolute magnitude of the theoretical and experimental curves, since the analytical result contains a global scale factor $I_{\mathrm{t}}$, which was fixated in Fig.~\ref{corr}A and B only to depict the variances.

The experimental covariances of the gains are also noisy (Fig.~\ref{corr}D). Both the theoretical and experimental covariances become significantly different from zero in those regions of the spectrum where the corresponding CMFs overlap. The theoretical result captures the sign of the (negative) covariance for the green-red interaction (yellowish curves) and the blue-green interaction (cyan), but not for the blue-red case (magenta). Yet, the analysis of Fig.~\ref{f:cf}B shows that the sign of this correlation is rather sensitive to the position of the wavelengths for which the cone fundamentals $q_i(\lambda)$ peak, and the sign inversion is compatible with the population dispersion of these positions \citep{Dartnall1983, Burns1993}.  

\section{Conclusions}
In summary, in this letter we used a Poisson model of photon absorption to predict and characterize the CMFs. In the first place, we deduced that the fraction of $L, M$ and $S$ cones composing the retina did not affect the shape of the trial-averaged CMFs (Eq.~\ref{e:efes}). We also analyzed the dependence of the CMFs with the primary colors used in the matching experiments, and searched for triplets of primaries that gave rise to particularly convenient CMFs. We later provided an analytical derivation of the variances and covariances of CMFs when the fluctuations are solely produced by Poisson noise in the photon absorption process. Quite notably, correlated CMFs were derived from uncorrelated photon absorption at $S, M$ and $L$ cones. Around $550$ nm, fluctuations along the direction $g_2 + g_3$ were accompanied by uncorrelated fluctuations along the direction $g_2 - g_3$. The relative frequency of $S, M$ and $L$ cones modified quantitatively but not qualitatively the variances and covariances, whereas the wavelengths $\lambda_i^{\mathrm{max}}$ at which $S, M$ and $L$ cones have maximal absorption probability produced noticeable changes in the variances and covariances. We then searched for primaries that minimized trial-to-trial variability, and concluded that the optimal triplet contained a short-wavelength primary $\lambda_1$ which coincided with $\lambda_{\mathrm{s}}^{\mathrm{max}}$, a middle-wavelength primary $\lambda_2$ that was shifted from $\lambda_{\mathrm{m}}^{\mathrm{max}}$ towards lower wavelengths in $7.6$ nm, and a long-wavelength primary $\lambda_3$ that was shifted from $\lambda_{\ell}^{\mathrm{max}}$ towards higher wavelengths in  $25.4$ nm. We were not able to find experimental data on intra-subject variability of CMFs obtained for monochromatic target stimuli, so we hope that the present study motivates psychophysical experiments. Given that the present study was based on the ideal-observer scheme implicit in the Cr\'amer-Rao bound, experiments characterizing the within-subject variability may or may not reveal the statistical properties described here. If future experiments confirm that they do, photoreceptor noise may be concluded to be a crucial ingredient in the perceptual variability of chromatic vision. Instead, if experiments happen to reveal a different behavior, the optimality assumed in subsequent processing stages cannot be ensured, and at least part of the observed variability must be associated to noisy processes downstream from photoreceptor capture. 

\section*{Acknowledgements}

This work was supported by Agencia Nacional de Investigaciones Científicas y Técnicas, Consejo Nacional de Investigaciones Científicas y Técnicas, Comisión Nacional de Energía Atómica and Universidad Nacional de Cuyo, all from Argentina.


\bibliographystyle{apalike}

\end{document}